\documentclass[prl,article,twocolumn,preprintnumbers,superscriptaddress,amsmath,amssymb,aps,longbibliography]{revtex4-1}

\usepackage{graphicx}
\usepackage{dcolumn}
\usepackage{bm}
\usepackage{multirow} 
\usepackage{comment}
\usepackage{url}
\usepackage{epsfig}
\usepackage{amsfonts}
\usepackage{amsmath}
\usepackage{bm}
\usepackage{graphicx}
\usepackage{color}
\usepackage{amssymb,amsmath,hyperref,slashed,color,graphicx,bm,cancel,url}
\usepackage[normalem]{ulem}
\usepackage[T1]{fontenc}
\usepackage{relsize}
\usepackage{gensymb}
\usepackage{fontawesome}
\usepackage[capitalize]{cleveref}
\usepackage[dvipsnames,table]{xcolor}

\hypersetup{colorlinks,
			citecolor      = niceblue, 
			linkcolor       = niceblue, 
			urlcolor        = niceblue, 
			anchorcolor = niceblue}
\definecolor{niceblue}{rgb}{0.1,0.2,0.6}

\definecolor{nicegreen}{rgb}{0., 0.75, 0.46}

\newcommand{\codeloc}{https://github.com/Harvard-Neutrino/MicroBooNE-analysis-2021}

\begin{document}

\preprint{CERN-TH-2021-195, FERMILAB-PUB-21-618-T, IPPP/21/50, FTPI-MINN-21-23}

\title{MicroBooNE and the \texorpdfstring{$\boldsymbol{\nu_e}$}{nu_e} Interpretation of the MiniBooNE Low-Energy Excess}

\author{C. A. Arg\"uelles}
\affiliation{Department of Physics \& Laboratory for Particle Physics and Cosmology, Harvard University, Cambridge, MA 02138, USA}

\author{I.~Esteban}
 \affiliation{Center for Cosmology and AstroParticle Physics
  (CCAPP), Ohio State University, Columbus, Ohio 43210, USA}
\affiliation{Department of Physics, Ohio State University, Columbus, Ohio 43210, USA}

\author{M.~Hostert}
\affiliation{Perimeter Institute for Theoretical Physics, Waterloo, ON N2J 2W9, Canada}
\affiliation{School of Physics and Astronomy, University of Minnesota, Minneapolis, MN 55455, USA}
\affiliation{William I. Fine Theoretical Physics Institute, School of Physics and Astronomy, University of
Minnesota, Minneapolis, MN 55455, USA}

\author{K.~J.~Kelly}
\affiliation{Theoretical Physics Department, CERN, Esplande des Particules, 1211 Geneva 23, Switzerland}

\author{J.~Kopp}
\affiliation{Theoretical Physics Department, CERN, Esplande des Particules, 1211 Geneva 23, Switzerland}
\affiliation{PRISMA+ Cluster of Excellence \& Mainz Institute for Theoretical Physics,
             Staudingerweg 7, 55128 Mainz, Germany}

\author{P.~A.~N.~Machado}
\affiliation{Particle Theory Department, Fermi National Accelerator Laboratory, Batavia, IL 60510, USA}

\author{I.~Martinez-Soler}
\affiliation{Department of Physics \& Laboratory for Particle Physics and Cosmology, Harvard University, Cambridge, MA 02138, USA}

\author{Y.~F.~Perez-Gonzalez}
\affiliation{Institute for Particle Physics Phenomenology, Durham University, South Road, Durham, United Kingdom}

\date{\today}

\begin{abstract}
{\centering{\href{\codeloc}{{\large\color{BlueViolet}\faGithub}}}
\\}
A new generation of neutrino experiments is testing the $4.8\sigma$ anomalous excess of electron-like events observed in MiniBooNE.
This is of huge importance for particle physics, astrophysics, and cosmology, not only because of the potential discovery of physics beyond the Standard Model, but also because the lessons we will learn about neutrino--nucleus interactions will be crucial for the worldwide neutrino program.
MicroBooNE has recently released results that appear to disfavor several explanations of the MiniBooNE anomaly.
Here, we show quantitatively that MicroBooNE results, while a promising start, unquestionably do not probe the full parameter space of sterile neutrino models hinted at by MiniBooNE and other data, nor do they probe the $\nu_e$ interpretation of the MiniBooNE excess in a model-independent way. 
Our analysis code is fully available in this \href{\codeloc}{GitHub repository}.
\end{abstract}

\maketitle

\emph{Introduction.---}
Sterile neutrinos have been postulated to explain various anomalies in neutrino experiments~\cite{Abazajian:2012ys}, in particular the excess of electron-neutrino events in the LSND~\cite{LSND:2001aii} and MiniBooNE~\cite{MiniBooNE:2020pnu} experiments.
If real, sterile neutrinos would revolutionize our understanding of early-universe cosmology~\cite{Dasgupta:2021ies}, modify neutrino emission from astrophysical sources~\cite{Arguelles:2019tum, Fiorillo:2020jvy, Fiorillo:2020zzj}, and force a re-evaluation of the Standard Model of particle physics.
Precision measurements in cosmology, astrophysics and particle physics will remain ambiguous until we clarify whether such sterile neutrinos exist. 

Recently, the MicroBooNE collaboration has released results scrutinizing the MiniBooNE low-energy excess (MBLEE). 
As the MicroBooNE detector is exposed to the same neutrino beam as MiniBooNE but has superior event reconstruction capabilities, it is able to differentiate between different interpretations of the MBLEE.
A first MicroBooNE analysis disfavors that the MBLEE is due to underestimated production of $\Delta$ baryons followed by decays to photons at a significance of 94.8\% C.L.~\cite{MicroBooNE:2021zai}.

Three distinct, complementary analyses have since been released, addressing whether the MBLEE is caused by an excess of electron-neutrinos in the beam~\cite{MicroBooNE:2021rmx, MicroBooNE:2021nxr, MicroBooNE:2021jwr, MicroBooNE:2021sne}.
These analyses compare the MicroBooNE data to a signal template defined by the assumption that the expected spectrum of the $\nu_e$ excess in MiniBooNE exactly matches the difference between the data and the best-fit background prediction.
Assuming this nominal template, the collaboration concludes that ``the results are found to be consistent with the nominal $\nu_e$ rate expectations from the Booster Neutrino Beam and no excess of $\nu_e$
events is observed''~\cite{MicroBooNE:2021rmx}.

However, as we show below, this approach is insufficient to exclude the $\nu_e$ interpretation of the MBLEE in a model-independent way, or even to exclude the sterile neutrino solution of the MBLEE.
It is important to explicitly address these questions, as they directly impact the future of the short-baseline neutrino program as well as a variety of alternative models~\cite{PalomaresRuiz:2005vf, Bai:2015ztj, Moss:2017pur, Dentler:2019dhz, deGouvea:2019qre, Hostert:2020oui} proposed to explain the MiniBooNE and LSND results. Previous attempts to constrain the LSND and MiniBooNE excesses suffered from insufficient sensitivity to cover the allowed parameter space~\cite{KARMEN:2002zcm, OPERA:2013wvp, ICARUS:2013cwr, NOMAD:2003mqg, Borodovsky:1992pn}, so the power of MicroBooNE must be quantified as we search for definitive answers to this 20-year-old puzzle. 
That is the goal of this letter.
We first analyze the constraints of MicroBooNE's latest results on $\nu_e$ appearance in MiniBooNE in a model-independent way, then we narrow our focus to sterile neutrinos. While doing this, we provide a methodology to analyze data that we hope will help in making future claims more robust. We make all analysis tools and results fully public in \cite{GitHubCode}.

For the first stage, we follow the MicroBooNE procedure: starting from a MiniBooNE event spectrum, we derive an expected excess of events in MicroBooNE, and we perform a statistical analysis of the data. 
After verifying that we reproduce MicroBooNE's results when using their nominal template, we repeat the analysis with a set of alternative templates that are equally successful at explaining the MBLEE.
These alternative templates are allowed due to the relatively large, nontrivial uncertainties in MiniBooNE.

For the second stage, we perform a fit of MicroBooNE data to a simple light sterile neutrino model.
We assume both a simplified oscillation scenario with only $\nu_\mu\to\nu_e$ appearance, as well as a fully-consistent oscillation model that accounts for oscillations in the MicroBooNE control samples and backgrounds.

\emph{Experimental Analysis.---}
To quantify the disagreement between MicroBooNE data and a $\nu_e$ interpretation of the MBLEE, we proceed as follows.
All our analyses start with a hypothesis for the MBLEE~\footnote{To directly compare with MicroBooNE results, we base our analysis on the 2018 MiniBooNE data~\cite{MiniBooNE:2018esg}.}.
To obtain the expected spectrum at MicroBooNE, we rescale the spectrum to account for the differences in exposure and detector mass between MiniBooNE and MicroBooNE~\cite{MicroBooNE:2021rmx}.
We then smear the events according to MicroBooNE's energy resolution~\cite{MicroBooNE:2021jwr, MicroBooNE:2021nxr}.
Finally, to infer MicroBooNE's energy-dependent $\nu_e$ detection efficiency, we apply all previous steps to the intrinsic $\nu_e$ background, and we choose the efficiency in each reconstructed energy bin such that our results match the official MicroBooNE background prediction~\cite{MicroBooNE:2021jwr, MicroBooNE:2021nxr}.
We have checked that our efficiencies are consistent with the energy-averaged efficiencies quoted in~\cite{MicroBooNE:2021jwr, MicroBooNE:2021nxr}, and that they generate an MBLEE prediction that matches the official MicroBooNE result. Our Supplemental Material provides more detail about checks and comparisons with the MicroBooNE expectations.

We focus on the Charged-Current Inclusive~\cite{MicroBooNE:2021nxr} and Quasielastic (CCQE)~\cite{MicroBooNE:2021jwr} channels,~\footnote{We note that the two samples are not statistically independent, as certain events will be present in both data sets. Therefore, the constraints that we derive from the two samples should be viewed as separate and should not combined as if they were independent.} as they comprise the leading statistical power of MicroBooNE, and we use the provided data releases wherever possible~\cite{DR_Inclusive,DR_CCQE}. 
For the Inclusive analysis, we have performed several statistical tests, including both Pearson-$\chi^2$ and CNP-$\chi^2$~\cite{Ji:2019yca}, as well as calculating a test statistic with or without deriving a constraint using the conditional covariance matrix formalism~\cite{MicroBooNE:2021nxr}.
For all tests, we find very good agreement with the results of~\cite{MicroBooNE:2021nxr}.
For clarity, in what follows we perform all statistical tests using the CNP-$\chi^2$ formalism with the full $(137,137)$ covariance matrix of~\cite{MicroBooNE:2021nxr,DR_Inclusive}.
For the CCQE analysis, in turn, we use a Poisson likelihood, where the expectation in each bin is treated as a nuisance parameter that is  constrained by the covariance matrix. 
Our test statistic in this latter analysis is then determined by profiling over these nuisance parameters.

\emph{Template results.---} Our first approach assesses whether MicroBooNE generically rules out a $\nu_e$ interpretation of the MBLEE.
In particular, the MiniBooNE signal has large, nontrivial, multiply-signed uncertainties.
These allow for different shapes of the MBLEE that could affect the prediction at MicroBooNE. 

We generate a set of MBLEE templates using a Markov-Chain Monte-Carlo (MCMC)~\cite{EMCEE} by independently re-scaling the normalization of the MiniBooNE backgrounds. Our template for the MBLEE is then given by the difference between the observed data and the re-scaled backgrounds~\footnote{We would like to stress that our goal is \emph{not} to reassess the MiniBooNE backgrounds, but to use them as a proxy to generate MBLEE shapes. Alternatively, we have also generated templates by independently varying the excess events in each bin.
Both approaches produce very similar results.}.
We group backgrounds in four classes: intrinsic $\nu_e$, mis-identified $\pi^0$, $\Delta\to\gamma$, and all others. 
To estimate how well each template fits the MiniBooNE data, we calculate its goodness-of-fit $p$-value using a $\chi^2$ test statistic with the MiniBooNE covariance matrix~\cite{MiniBooNE:2021bgc}.
We follow the MiniBooNE prescription, that the test-statistic follows a $\chi^2$ distribution with 8.7 degrees of freedom (while our template model has 4 d.o.f.).

This generates a set of MBLEE templates compatible with MiniBooNE. To translate them to MicroBooNE, we have to generate an MBLEE template \emph{before} MiniBooNE detector effects.
For this, we apply the D'Agostini iterative unfolding method~\cite{DAgostini:1994fjx,muBooNE-NOTE-1043}, whose validity we have checked for a variety of spectra (see Supplemental Material). We do not assign any uncertainty to this procedure, as the purpose of unfolding is just to inspire some choices of MBLEE spectra and see if some are compatible with MiniBooNE and MicroBooNE. 

\begin{figure}[t]
\centering
\includegraphics[width=0.49\textwidth]{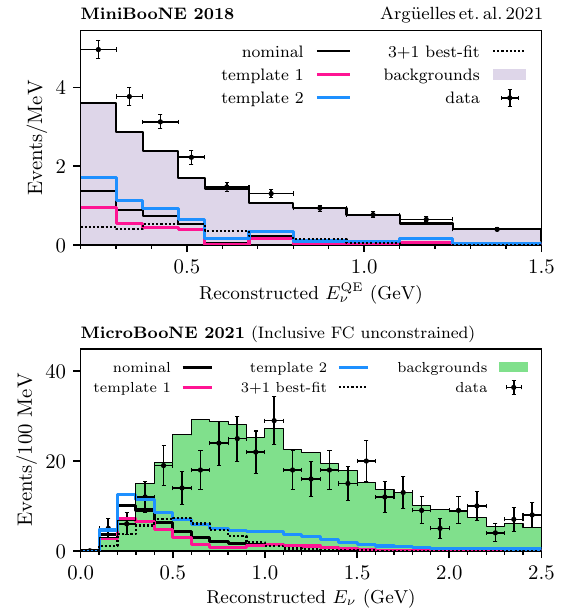}
\caption{
Event rate at MiniBooNE (top) and MicroBooNE (bottom) as a function of reconstructed neutrino energy.
We show several MiniBooNE templates, including that of the 3+1 oscillation best-fit, as non-stacked histograms. The bottom panel shows the spectra predicted by these templates in the MicroBooNE Inclusive fully contained channel.
\label{fig:nue_excess_shape}
}
\end{figure}

\Cref{fig:nue_excess_shape} demonstrates our procedure.
We show three different MiniBooNE templates (upper panel) and the corresponding predicted MicroBooNE excesses (lower panel): the nominal template given by the difference between the observed data and the best-fit background estimate (solid black); a template with significantly more events at low energies but a $p$-value of $87\%$ (blue); a template with fewer events and a $p$-value of $87\%$ (red); and a template given by the best-fit $\nu_\mu\to\nu_e$ two-neutrino oscillations, corresponding to $\Delta m^2=0.041$~eV$^2$ and $\sin^2 2\theta= 0.92$, which has a $p$-value of $20\%$ (dotted black).

As we see, MiniBooNE uncertainties allow for very different shapes and rates of the MBLEE (including those coming from the sterile neutrino hypothesis) to provide a good fit to the data.
These generate different predictions at MicroBooNE that will be excluded with a different statistical significance.
This is crucial in constraining the interpretation of the MBLEE in terms of $\nu_e$ events at MicroBooNE.

For the template analysis results, we focus on the Inclusive channel, which provides the best constraints on the $\nu_e$ interpretation of the MBLEE.
We organize the templates in three categories by decreasing goodness-of-fit, $p > $ $80\%$, $10\%$, and $1\%$; and classify them by their signal strength, defined as $N/N_\mathrm{LEE}$ with $N$ the number of excess events that the template predicts at MiniBooNE and $N_\mathrm{LEE}=360$ the observed number of excess events.

\begin{figure}[t]
\centering
\includegraphics[width=0.49\textwidth]{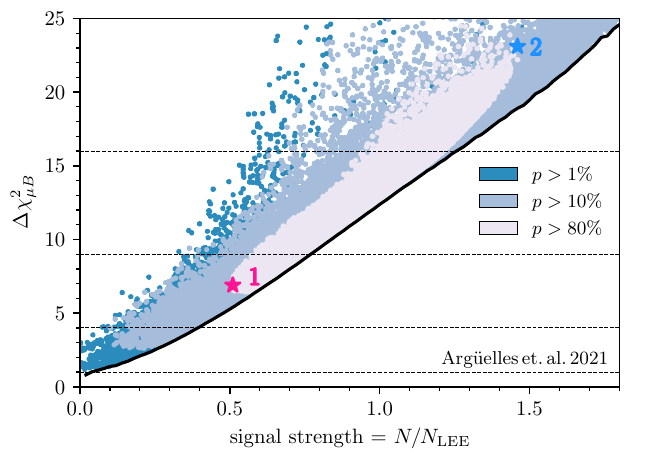}
\caption{$\Delta \chi^2$ of the MicroBooNE Inclusive analysis with respect to the no-excess hypothesis, for various templates found by our MCMC.
Each point corresponds to a specific template that provides a good fit to MiniBooNE data with a $p$-value greater than $80\%$, $10\%$, and $1\%$ (shades of blue).
The stars correspond to templates 1 and 2 presented in Fig.~\ref{fig:nue_excess_shape}. \label{fig:signal_strength}}
\end{figure}

\Cref{fig:signal_strength} shows the result of the template analysis.
Each point corresponds to a different template, colored according to the three goodness-of-fit categories defined above. 
For each template, we compute the corresponding MicroBooNE $\chi^2_{\mu \rm B}$, and construct the difference with respect to the no-excess hypothesis, $\Delta \chi^2_{\mu \rm B}$.
We also show as a black line the result of profiling over all templates with the same signal strength.
The horizontal lines then correspond to the MicroBooNE $1$, $2$, $3$, and $4\sigma$ exclusion limits~\cite{wilks1938,Algeri:2019lah}.

As we see in Fig.~\ref{fig:signal_strength}, introducing shape and normalization uncertainties in the MBLEE template can either enhance or mitigate MicroBooNE's sensitivity.
To illustrate the variability of the template shapes and normalizations, we have marked with a star the two templates shown in \cref{fig:nue_excess_shape}, corresponding to two extreme points in the $p > 80\%$ region.

Many templates that are a good fit to MiniBooNE data cannot be excluded by MicroBooNE --- we observe a large number of templates with good fits to MiniBooNE data, $p > 80\%$ ($10\%$), well below the $\Delta \chi^2_{\rm \mu B} = 9$ $(4)$ line.
We thus conclude that, while recent MicroBooNE results indeed constrain the $\nu_e$ interpretation of the MiniBooNE excess in a model-independent way, they do not completely rule it out. Because of the correlated systematic uncertainties between MiniBooNE and MicroBooNE, to fully establish the compatibility of these templates, a joint analysis is required.

\emph{Sterile Neutrino Analysis.---}
The analysis above does not rely on any specific particle physics model.
As an example of a physics model that can explain the MBLEE, we turn to light sterile neutrinos.
They provide a simple scenario that could lead to $\nu_\mu\to\nu_e$ transitions at short baselines, and have been extensively studied in the literature~\cite{Abazajian:2012ys,Gariazzo:2015rra,Diaz:2019fwt,Boser:2019rta,Dasgupta:2021ies}. Here, we do not rely on the unfolding technique discussed above, as we simulate expected distributions in MiniBooNE with respect to the true neutrino energy.

To perform analyses including sterile neutrinos, we use~\cite{Dentler:2018sju} and first calculate, as a function of oscillation parameters, the expected MBLEE.
Using the same procedure discussed above, we map these spectra into the expected excesses in MicroBooNE's Inclusive and CCQE analyses. Leveraging~\cite{DR_Inclusive,DR_CCQE}, we can also account for oscillations of the $\nu_\mu$ and $\nu_e$ Charged-Current (CC) background expectations in MicroBooNE's analyses to allow for a complete, four-neutrino oscillation analysis~\footnote{More concretely, we use the data releases provided along with~\cite{MicroBooNE:2021nxr,DR_Inclusive} to determine the expected $\nu_{\mu,e}$ CC fully- and partially-contained spectra as a function of \textit{true} neutrino energy. Oscillations are included with respect to true energy, then the distributions are mapped into \textit{reconstructed} neutrino energy (again, using~\cite{DR_Inclusive}) where test statistics are calculated.}. 

We start by discussing the results of the simplified sterile neutrino model, which assumes the backgrounds to be independent of the sterile neutrino parameters.
This simplified model is parametrized by a squared-mass difference $\Delta m^2_{41}$ and an effective mixing angle $\sin^2 2\theta_{\mu e}\equiv 4|U_{e 4}U_{\mu4}|^2$ with $U$ the leptonic mixing matrix.

\begin{figure}[t]
\centering
\includegraphics[width=0.49\textwidth]{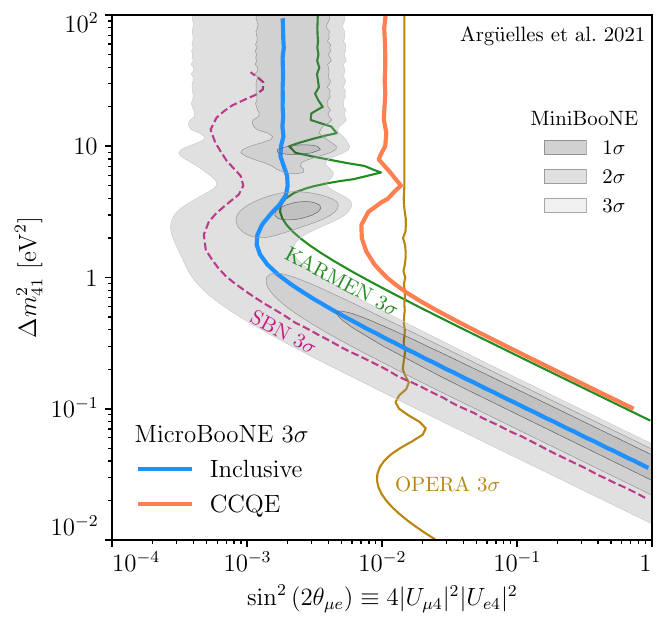}
\caption{\label{fig:sterile_constraint}
MicroBooNE constraints on sterile neutrino parameter space at 3$\sigma$ C.L. (blue, Inclusive and orange, CCQE). 
For reference, we show the MiniBooNE $1$-, $2$-, and $3$-$\sigma$ preferred regions in shades of grey~\cite{MiniBooNE:2018esg}, the future sensitivity of the three SBN detectors (pink)~\cite{Machado:2019oxb}, and existing constraints from KARMEN (green)~\cite{KARMEN:2002zcm} and OPERA (gold)~\cite{OPERA:2018ksq}.
}
\end{figure}

\Cref{fig:sterile_constraint} presents the results of our analyses of MicroBooNE's Inclusive and CCQE channels in blue and orange, respectively, at 3$\sigma$ C.L. We first note that the Inclusive analysis has more constraining power than the CCQE analysis. This can be traced back to the detection efficiencies, which for the MiniBooNE and MicroBooNE CCQE analyses decrease at large energies, while in the Inclusive MicroBooNE analysis they stay constant. As sterile neutrinos predict a non-negligible excess at high energies, the Inclusive analysis is more powerful.

As we see from \cref{fig:sterile_constraint}, MicroBooNE data, at $3\sigma$ CL, disfavor part of the region preferred by MiniBooNE at the same CL. Nevertheless, we find that there is still a large viable fraction of the parameter space, even within $1\sigma$ CL preferred region of MiniBooNE.
We find it unlikely that future MicroBooNE results will significantly improve on this, even though MicroBooNE has only analyzed about half of their data set, because of a deficit in their Inclusive data that generates more sensitivity than expected (\textit{cf.} \cref{fig:nue_excess_shape} and the Supplemental Material; this could be due to an underfluctuation in the data or to background mismodeling).
This highlights the importance of searching for sterile neutrinos with the three SBN detectors --- SBND, MicroBooNE, and ICARUS --- which will probe the full $2\sigma$ region preferred by MiniBooNE with less dependence on the neutrino cross section and flux.

Finally, we stress that a fully-consistent four-neutrino analysis should also consider oscillations of the backgrounds. 
This is relevant at MiniBooNE~\cite{Kopp:2013vaa,Dentler:2019dhz}, and even more for the MicroBooNE Inclusive analysis: while the former has large non-neutrino induced backgrounds, the dominant background in the latter is beam-$\nu_e$ contamination.
Moreover, since other neutrino samples (particularly CC $\nu_\mu$) are used to constrain systematics and backgrounds, oscillations should also be considered for these samples.
\begin{figure}[t]
\centering
\includegraphics[width=0.49\textwidth]{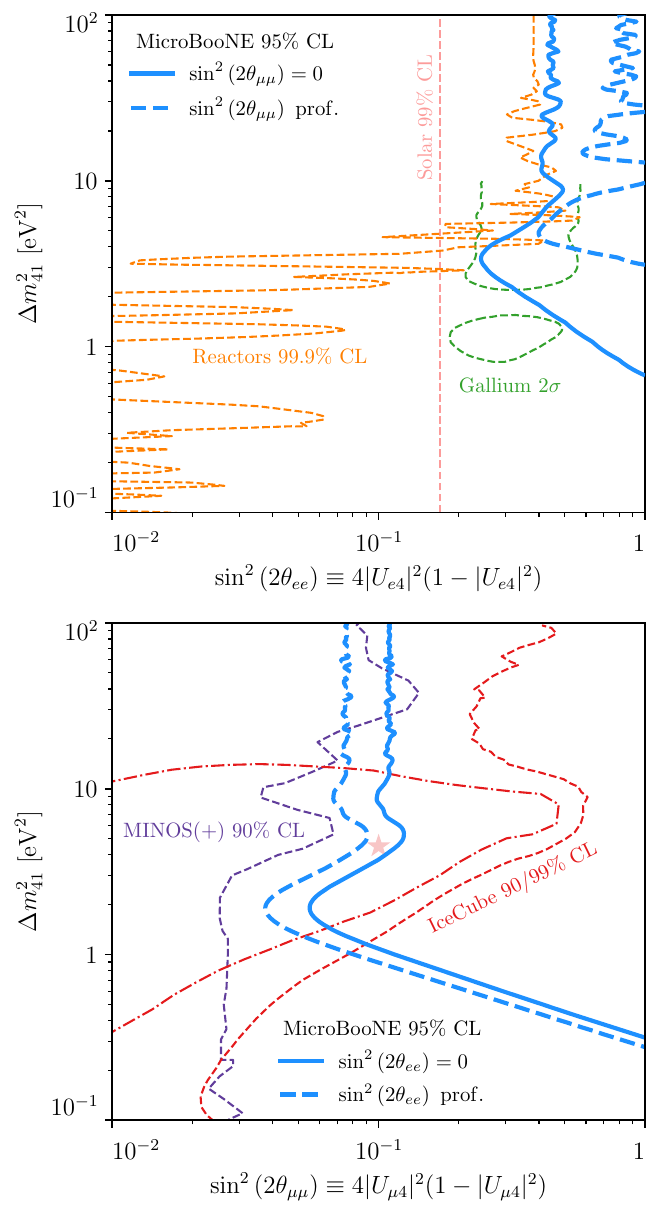}
\caption{\label{fig:full_oscillation}
MicroBooNE constraints on $\Delta m_{41}^2$ and $\sin^2\left(2\theta_{ee}\right)$ (left) or $\sin^2\left(2\theta_{\mu\mu}\right)$ (right). In each panel, we have either fixed (solid lines) or profiled over (dashed) the unshown mixing angle. For comparison, we show existing constraints and preferred regions (see~\cite{Barinov:2021asz,SAGE:1998fvr,Abdurashitov:2005tb,GNO:2005bds,Kaether:2010ag,Goldhagen:2021kxe,Berryman:2020agd,Declais:1994su,DANSS:2018fnn,DayaBay:2018yms,DoubleChooz:2019qbj,NEOS:2016wee,RENO:2018dro,MINOS:2017cae,IceCube:2020phf}).}
\end{figure}

Figure~\ref{fig:full_oscillation} presents our results in a consistent four-neutrino approach, considering oscillations of all $\nu_e$ and $\nu_\mu$ samples.
We show the MicroBooNE-Inclusive 95\% CL constraints on $\Delta m^2_{41}$ and $\sin^2\left(2\theta_{ee}\right) \equiv 4 |U_{e 4}|^2 (1-|U_{e 4}|)^2$ (top panel) or $\sin^2\left(2\theta_{\mu\mu}\right) \equiv 4 |U_{\mu 4}|^2 (1-|U_{\mu 4}|)^2$ (bottom panel). In each panel of Fig.~\ref{fig:full_oscillation} we perform two analyses (both in blue): solid lines present the constraint on a mixing angle when the other is fixed to zero, whereas dashed lines present the constraint when we profile over the other mixing angle.
The disappearance prospects for $\nu_e$ are compared against hints of sterile neutrinos in Gallium experiments~\cite{SAGE:1998fvr,Abdurashitov:2005tb,GNO:2005bds,Kaether:2010ag,Barinov:2021asz} and constraints from solar~\cite{Goldhagen:2021kxe}, and reactor antineutrino~\cite{Berryman:2020agd,Declais:1994su,DANSS:2018fnn,DayaBay:2018yms,DoubleChooz:2019qbj,NEOS:2016wee,RENO:2018dro} experiments.
The bottom panel, showing MicroBooNE $\nu_\mu$ disappearance constraints, is contrasted against constraints from MINOS/MINOS+~\cite{MINOS:2017cae} and results from IceCube~\cite{IceCube:2020phf}, including a 90\% C.L. preferred region and a best-fit point. For further justification of this test-statistic and coverage studies for both Figs.~\ref{fig:sterile_constraint} and~\ref{fig:full_oscillation}, see the Supplemental Material.

As we see, even in the absence of neutrino appearance, \textit{i.e.}, $U_{e4}$ or $U_{\mu4}$ equal zero, MicroBooNE can still set a limit on neutrino disappearance.
For muon neutrinos, the disappearance sensitivity comes from the large $\nu_\mu$ data sample.
For electron neutrinos, in turn, the sensitivity derives from the large $\nu_e$ background.

\begin{figure}[t]
\centering
\includegraphics[width=0.45\textwidth]{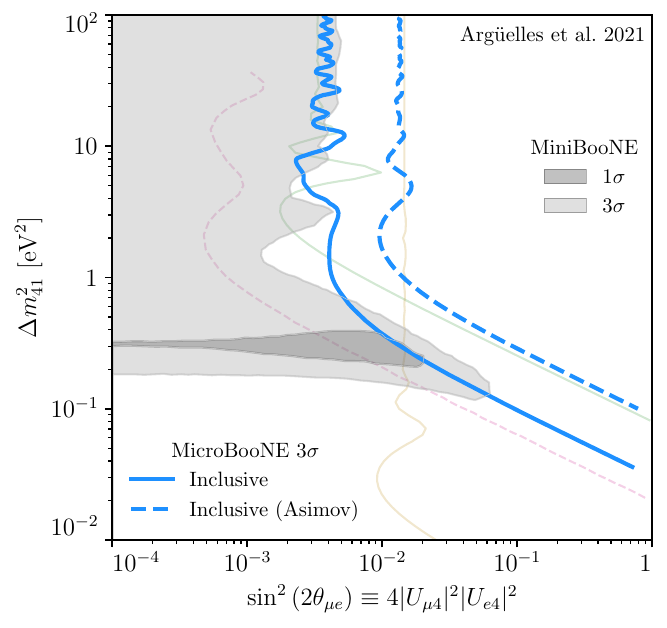}
\caption{\label{fig:full_oscillation_appearance}
MicroBooNE constraint on $\Delta m_{41}^2$ and $\sin^2\left(2\theta_{\mu e}\right)$ after profiling over the unshown mixing angle in a consistent four-neutrino analysis. Preferred MiniBooNE regions~\cite{Brdar:2021cgb} are shown in grey. Other constraints/projections, faded for clarity, are identical to those in Fig.~\ref{fig:sterile_constraint}.}
\end{figure}

As one final remark on the importance of the complete four-neutrino analysis, Fig.~\ref{fig:full_oscillation_appearance} shows the $3\sigma$ CL constraint from MicroBooNE-Inclusive as a function of $\sin^2\left(2\theta_{\mu e}\right)$ after profiling over $\sin^2\left(2\theta_{\mu\mu}\right)$ in comparison with $1\sigma$ and $3\sigma$ CL preferred regions of MiniBooNE under the same set of assumptions~\cite{Brdar:2021cgb}.
Even more than in Fig.~\ref{fig:sterile_constraint}, we see that $3\sigma$-CL-allowed MiniBooNE parameter space persists despite the MicroBooNE-Inclusive constraints.

Our results emphasize that, while the signal-oscillation-only analysis is simple and intuitive, accounting for oscillations in \textit{all} samples is the only fully-consistent approach and can affect the interpretation of the results.
This will be even more relevant for the full SBN program, particularly due to different oscillation effects among the three detectors as well as due to the increased analysis sensitivity. We \textit{strongly} advocate for the adoption of this standard moving forward with short-baseline searches for anomalous neutrino (dis)appearance.

In this complete picture, we find results consistent with no oscillations, with a best-fit point at $\Delta m_{41}^2 = 1.38$ eV$^2$, $\sin^2\left(2\theta_{ee}\right) = 0.2$, and $\sin^2\left(2\theta_{\mu\mu}\right) = 0$ with a significance of $0.4\sigma$.
Our results do not agree with~\cite{Denton:2021czb}.
We believe that this stems from the treatment of systematic uncertainties, as we consider correlated systematics by using~\cite{DR_Inclusive}; and due to the fact that we account for oscillations in the partially-contained $\nu_e$ sample, which is used to obtain the constrained fully-contained $\nu_e$ sample.
We also implement oscillations as a function of true neutrino energy.

\emph{Conclusions.---} 
Does MicroBooNE rule out the $\nu_e$ interpretation of the MiniBooNE low-energy excess?
And does it disfavor the sterile neutrino explanation of the excess? 
While current MicroBooNE analyses give us invaluable insights on the MiniBooNE anomaly, we find that they still do not provide definitive answers to either of these two questions.
Uncertainties on MiniBooNE backgrounds significantly impact MicroBooNE's reach, and consequently, the MiniBooNE puzzle remains wide open.
To demonstrate this quantitatively, we have developed a model-independent analysis and we have carried out a fully-consistent sterile neutrino fit of MicroBooNE data in the context of the MiniBooNE excess.
In the first analysis, we find MiniBooNE excess spectra with goodness-of-fit better than $10\%$ that are allowed by MicroBooNE data at $<2\sigma$.
In the sterile neutrino analysis, we find that MicroBooNE's $3\sigma$ exclusion does not cover the entire MiniBooNE LEE allowed region.

Our findings highlight the importance of running the full SBN program, and of complementing it with the worldwide efforts to search for light sterile neutrinos in reactor~\cite{STEREO:2018blj,PROSPECT:2018dnc,Alekseev:2016llm,Serebrov:2020kmd,Licciardi:2021hyi}, radioactive source~\cite{Barinov:2021asz}, accelerator~\cite{Ajimura:2017fld,Suekane:2020avt,Alonso:2021kyu,Alonso:2021jxx}, solar~\cite{Goldhagen:2021kxe,deGouvea:2021ymm}, and atmospheric neutrino~\cite{Super-Kamiokande:2014ndf,ANTARES:2018rtf,IceCube:2020phf,Smithers:2021orb,Domi:2021cix,Wang:2021gox} experiments.
Together, these experiments will have sufficient sensitivity to answer this decades-old puzzle once and for all.

\begin{acknowledgments}
\emph{Acknowledgements.---}
We are particularly grateful to John Beacom and Steven Prohira for invaluable discussions and involvement in the early stages of this work. We thank Jeff Berryman and Bryce Littlejohn for discussions regarding reactor antineutrino measurements.
P.M. is  grateful to several members of CCAPP at Ohio State University for many discussions on the MicroBooNE results.
Fermilab is operated by the Fermi Research Alliance, LLC under contract No. DE-AC02-07CH11359 with the United States Department of Energy. This project has received support from the European Union's Horizon 2020 research and innovation program under the Marie Sk\l{}odowska-Curie grant agreement No. 860881-HIDDeN.
C.A.A. is supported by the Faculty of Arts and Sciences of Harvard University, and the Alfred P.  Sloan Foundation. IMS is supported by the Faculty of Arts and Sciences of Harvard University.  Perimeter Institute is supported by the Government of Canada through the Department of Innovation, Science and Economic Development and by the Province of Ontario through the Ministry of Research, Innovation and Science.
\end{acknowledgments}

\bibliographystyle{apsrev4-1}
\bibliography{references}

\clearpage

\onecolumngrid
\appendix

\ifx \standalonesupplemental\undefined
\setcounter{page}{1}
\setcounter{figure}{0}
\setcounter{table}{0}
\setcounter{equation}{0}
\fi

\renewcommand{\thepage}{Supplementary Methods and Tables -- S\arabic{page}}
\renewcommand{\figurename}{SUPPL. FIG.}
\renewcommand{\tablename}{SUPPL. TABLE}

\renewcommand{\theequation}{A\arabic{equation}}

\section{Comparison of Results to Asimov Sensitivity Expectations}
In the \emph{Sterile Neutrino Analysis} section, see Fig. 3 of the main text, we presented the $3\sigma$  constraints on $\sin^2\left(2\theta_{\mu e}\right)$ vs. $\Delta m_{41}^2$ from the MicroBooNE Inclusive and CCQE analyses, and contrasted these constraints against existing and proposed future results. For completeness, in this appendix we provide a comparison between the results from MicroBooNE data and the expected Asimov sensitivity, i.e., the constraints obtained by replacing the observed data with the expected background.

\begin{figure}[h]
\centering
\includegraphics[width=0.45\textwidth]{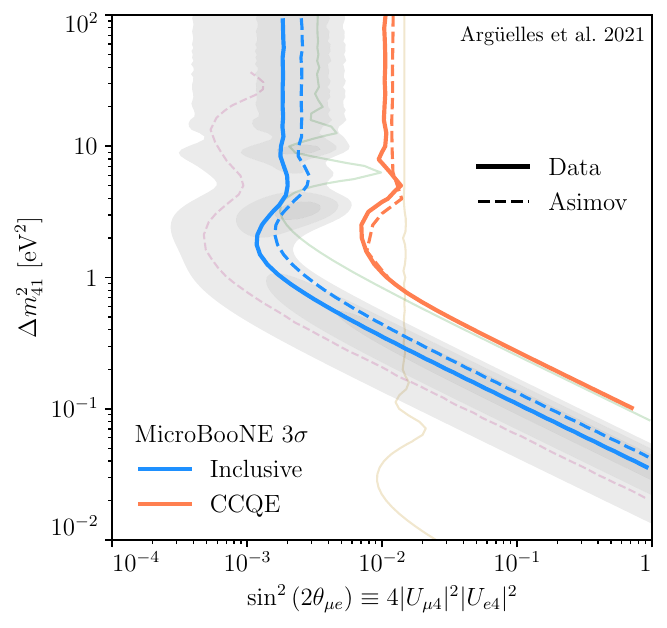}
\caption{\label{fig:sterile_constraint_asimov}
Comparison of MicroBooNE constraints (solid lines) with the expected Asimov sensitivity (dashed) of each analysis, both at $3\sigma$ C.L. Blue lines correspond to the Inclusive analysis, and orange to the CCQE analysis. Constraints/preferred regions in this panel, dimmed for clarity, are identical to those shown and labeled in Fig.~3 of the main text.
}
\end{figure}
Suppl. Fig.~\ref{fig:sterile_constraint_asimov} presents this comparison. As discussed in the main text,  the deficit of observed events relative to the expected background rate leads to a more powerful constraint than expected in both analyses, but especially so in the Inclusive one (see e.g.~Fig.~1 of the main text).

\begin{figure}[t]
\centering
\includegraphics[width=\textwidth]{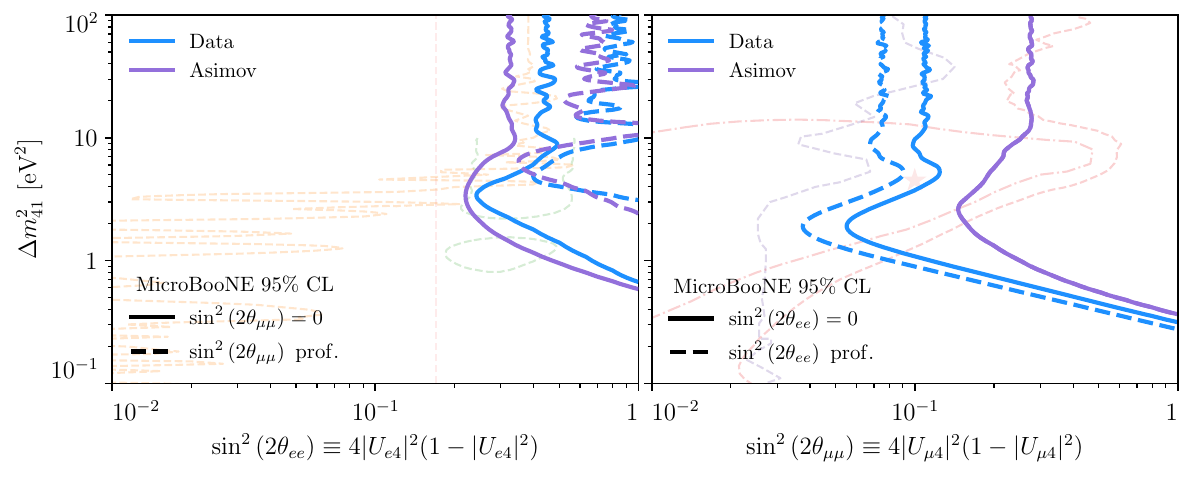}
\caption{\label{fig:full_oscillation_asimov}
Comparison of MicroBooNE constraints (blue) with the Asimov sensitivity expectation (purple) at 95\% C.L. as a function of $\sin^2\left(2\theta_{ee}\right)$ and $\sin^2\left(2\theta_{\mu\mu}\right)$ vs. $\Delta m_{41}^2$. Faded lines are identical to those in Fig.~4 of the main text and are only dimmed for clarity of comparison. In each panel, the solid (dashed) lines correspond to fixing (profiling over) the unseen mixing angle. The Asimov-expected sensitivity in the right panel for $\sin^2\left(2\theta_{ee}\right) = 0$ and when it is profiled are identical, so only the solid purple line is visible.
}
\end{figure}

Suppl. Fig.~\ref{fig:full_oscillation_asimov} repeats this data/Asimov comparison for the results of Fig.~4 of the main text, comparing data results in blue with Asimov expectations in purple. 
Here we focus, as in the main text, on the MicroBooNE Inclusive analysis. In both panels, as in Fig.~4 of the main text, solid lines correspond to the analysis with the other mixing angle fixed to zero, whereas the dashed lines include profiling over the unseen angle.
We see in the right panel that the data result exceeds the sensitivity greatly -- this is driven by the observed $\nu_\mu$ CC excess observed in both the Fully-Contained and Partially-Contained samples. These samples also modify the expectation of $\nu_e$ CC event rates due to large correlations, which has a non-negligible impact on the sensitivity to $\sin^2\left(2\theta_{ee}\right)$ in the left panel. In the right panel of Fig.~\ref{fig:full_oscillation_asimov}, there is no difference in the Asimov-expected sensitivity between $\sin^2\left(2\theta_{ee}\right) = 0$ and when profiling over this angle, so there is no distinct dashed purple line.

\section{Unfolding Cross-checks}

\begin{figure}[t]
\centering
\includegraphics[width=0.5\textwidth]{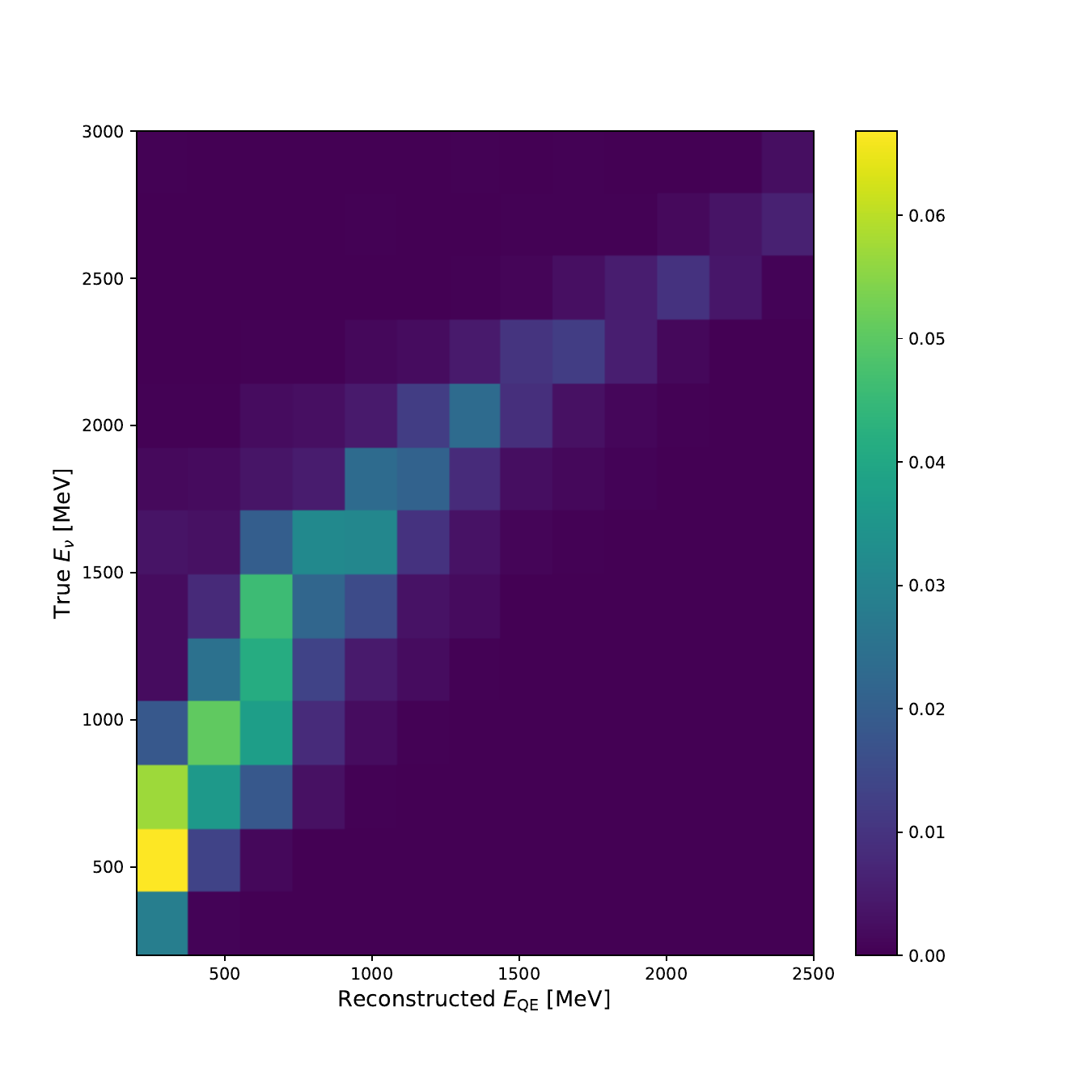}
\caption{\label{fig:mB_Res_M} Response Matrix obtained from MiniBooNE's 2018 data release, used in our unfolding procedure.}
\end{figure}

In this section, we present some of the checks performed to the unfolding process used for the template analysis. 
As mentioned in the main text, we follow the MicroBooNE procedure established in~\cite{muBooNE-NOTE-1043}. 
The main ingredient for the unfolding is the MiniBooNE response matrix. 
We obtain such a matrix using their 2018 data release~\cite{Aguilar-Arevalo:2019owf} and their established method~\cite{MiniBooNE-DR-2010}. 
For completeness, we present the response matrix in the Suppl.~Fig.~\ref{fig:mB_Res_M}.

\begin{figure}[t]
\centering
\includegraphics[width=0.5\textwidth]{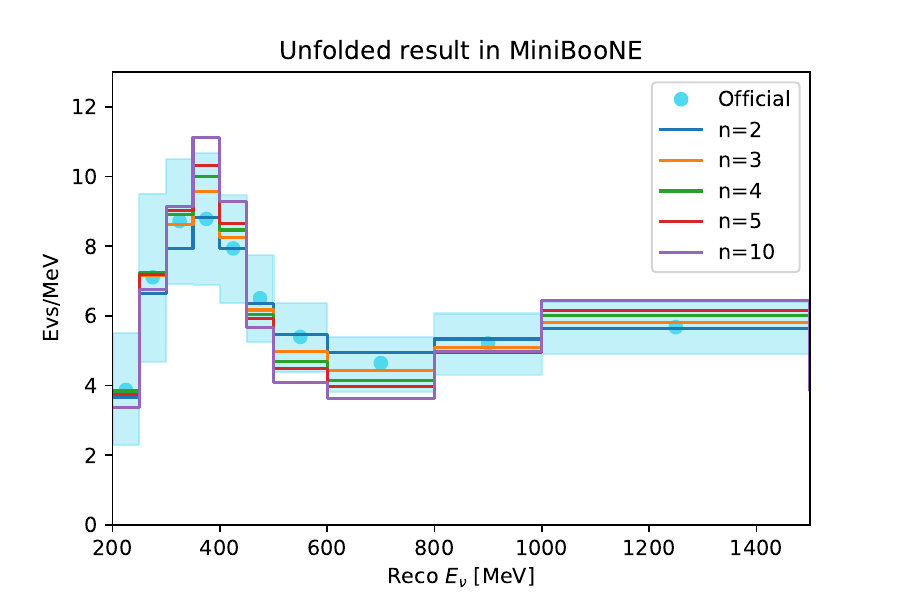}
\caption{\label{fig:Unf_n} Dependence of the unfolded excess events on the number of steps used in the D'Agostini method. We present in light-blue the official MicroBooNE unfolding together with their uncertainty. }
\end{figure}

To scrutinize the independence of our results on the Unfolding procedure, we have varied the number of steps used in the D'Agostini method to establish the reliance of the unfolded spectrum on such number. 
We present in Supp.~Fig.~\ref{fig:Unf_n} the unfolded excess events in MiniBooNE obtained for different values of the number of steps. 
We also show the MicroBooNE official unfolding, obtained from Ref.~\cite{muBooNE-NOTE-1043}. 
We observe that, for small values of the number of steps ($n<5$), our unfolded spectrum lies within the official range from MicroBooNE, and are close to the central values obtained by the Collaboration. 
Meanwhile, if the number of steps is larger, for instance $n=10$ in the figure, the unfolding gets closer to inverting the response matrix, which yields large variances.  

\begin{figure}[t]
\centering
\includegraphics[width=0.5\textwidth]{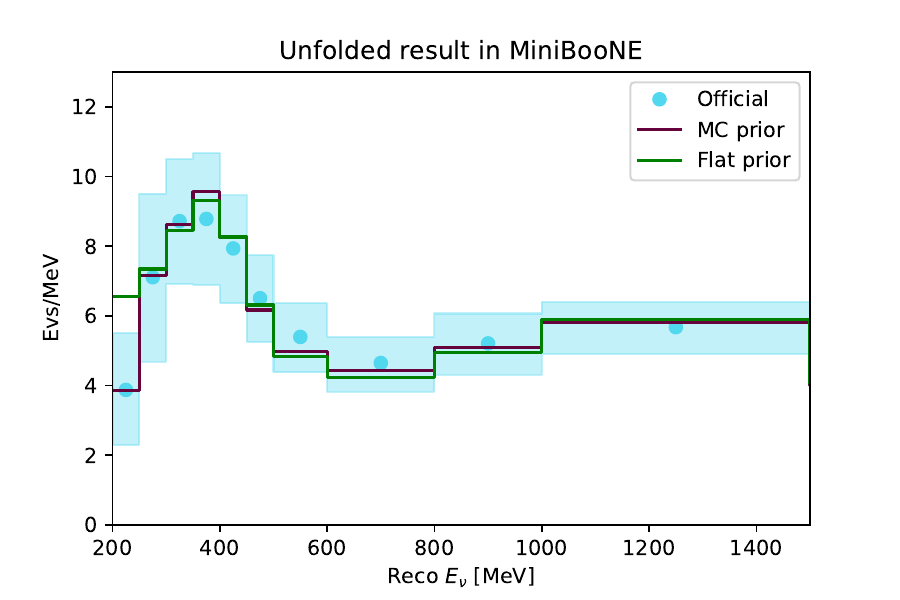}
\caption{\label{fig:Unf_p} Our unfolded excess considering as priors the MC intrinsic $\nu_e$ events (burgundy) and a flat prior in energy (green). We present in light-blue the official MicroBooNE unfolding together with their uncertainty. }
\end{figure}

For the previous analysis we used as prior the MC predicted $\nu_e$ events assuming no excess present in the MiniBooNE data. 
As an additional validation, we have considered a prior flat on the energy for the unfolding. 
The unfolded spectra obtained assuming the MC (burgundy) and flat (green) priors are presented in the Supp.~Fig.~\ref{fig:Unf_p}. 
We observe a significant discrepancy between the unfolded spectra only in the first bin, which nevertheless does not affect our results.

\begin{figure}[t]
\centering
\includegraphics[width=0.75\textwidth]{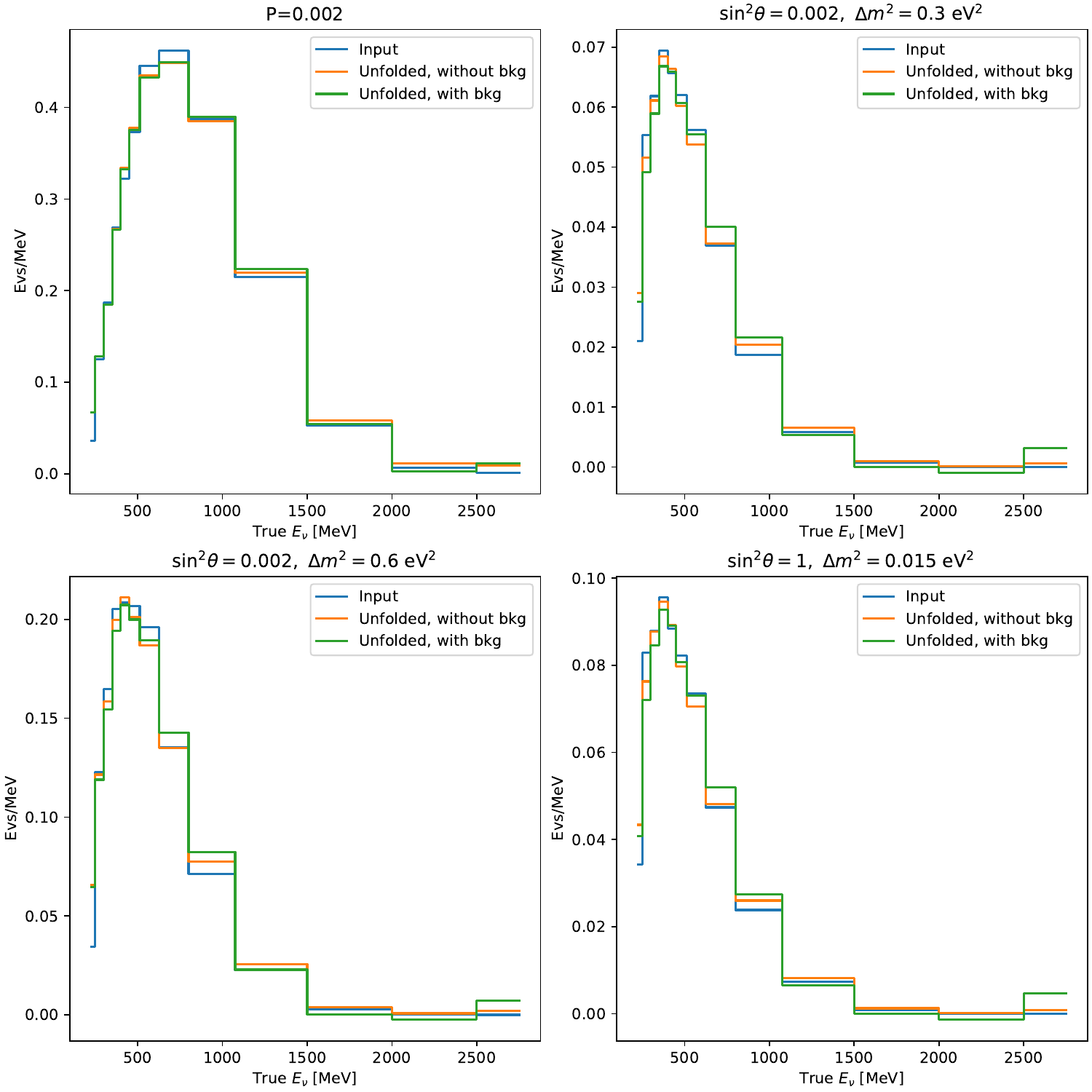}
\caption{\label{fig:Compar_1} Input (blue) and unfolded (orange and green) spectra assuming a $\nu_\mu\to\nu_e$ oscillation driven by a sterile state with an energy independent probability $P=0.002$ (top-left), and taking $(\sin^2\theta,\Delta m^2)=\{(0.002,0.3~{\rm eV^2}),(0.002,0.6~{\rm eV^2}),(1,0.015~{\rm eV^2})\}$, the top-right, bottom-left and bottom-right panels, respectively. The unfolded spectrum is obtained by determining the oscillation spectrum in terms of reconstructed energy and the unfolding the resulting events using the D'Agostini method. In the ``folding-unfolding'' procedure, we considered two distinct situations were we included (green) or not (orange) the MC simulated backgrounds.}
\end{figure}

As a further validation, we investigated the robustness of the unfolding procedure by artificially creating spectra in true energy, which were translated to events in reconstructed energies using the MiniBooNE response matrix, and then unfolded back to true energy. 
With this procedure, we intended to determine the stability of the unfolding.
We considered as initial spectra the events obtained by assuming $\nu_\mu\to\nu_e$ oscillations generated by a sterile neutrino, as prescribed in~\cite{MiniBooNE-DR-2010}. 
For this test, we use $n=3$ as the number of steps in the D'Agostini method, the MC intrinsic $\nu_e$ spectra as prior, and we use the same binning as the collaboration.
In Supp.~Fig.~\ref{fig:Compar_1}, we present the input (blue) and the unfolded without any additional background events (orange) spectra assuming some specific parameters of the oscillation probability, $P=0.002$, a constant probability (top left), $(\sin^2\theta,\Delta m^2)=\{(0.002,0.3~{\rm eV^2}),(0.002,0.6~{\rm eV^2}),(1,0.015~{\rm eV^2})\}$ in the top-right, bottom-left and bottom-right panels, respectively. 
From this test, we observe that the ``folding-unfolding'' yields quite acceptable results, even if they do not exactly reproduce the input spectra.
Certainly, this is expected since the unfolding is a information-losing procedure \emph{per se}.

The previous test spectra only included events coming from the assumed oscillations, that is, it did not contain any background events. 
To understand if the presence of a such background events would alter the unfolding of the signal, we included the MC simulated background from the MiniBooNE collaboration, and then repeated the previously mentioned method.
After unfolding back to the true neutrino energy, we subtracted the background, which is assumed to be known.
The green lines in the Supp.~Fig~\ref{fig:Compar_1} correspond to the unfolded spectrum containing the background.
In the case where the oscillation probability is energy independent, we observe that in both scenarios, with and without background, the unfolded spectra are rather similar, although differing from the original spectrum.
When there are pronounced features in the input true energy spectrum, we observe larger discrepancies between the unfolded spectra. 

\begin{figure}[t]
\centering
\includegraphics[width=0.6\textwidth]{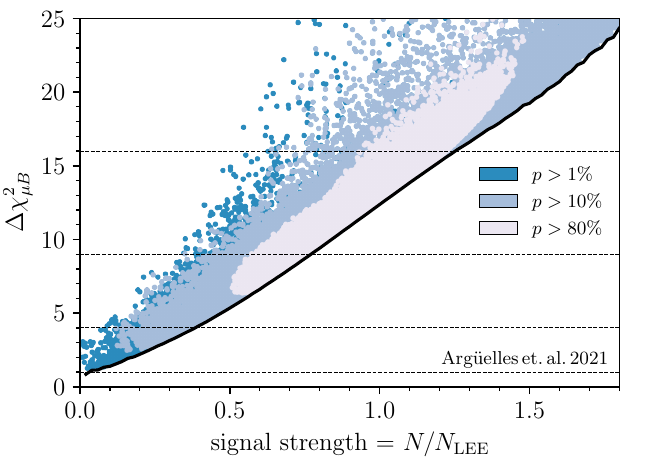}
\caption{\label{fig:Compar_2} MicroBooNE sensitivity to the MBLEE template. In the statistical analysis, we have included the shape and normalization uncertainties over the templates. For the unfolding procedure, we follow the D'Agostini method with 5 steps.}
\end{figure}

Finally, we explore the impact of the number of steps considered in the unfolding procedure on the MicroBooNE’s sensitivity to the MBLEE. Following the D'Agostini method and 5 steps, we consider a set of alternative templates allowed by the shape and normalization uncertainties in MiniBooNE, Fig.~\ref{fig:Compar_2}. We find no difference in MicroBooNE's sensitivity to the templates by changing the number of steps in the unfolding.

\section{Test Statistic Coverage Checks}
In this appendix, we describe some verification of the test statistic that is used in deriving confidence levels throughout our different analyses. Specifically, we verify that the $\Delta\chi^2$ used to determine the MicroBooNE-Inclusive excluded regions follows the chi-squared distribution for two degrees of freedom in the analysis of Fig.~3 of the main text. We perform this check for two specific points in parameter space, (a) $\sin^2\left(2\theta_{\mu e}\right) = 8 \times 10^{-4}$, $\Delta m_{41}^2 = 2.2$ eV$^2$ and (b) $\sin^2\left(2\theta_{\mu e}\right) = 10^{-1}$, $\Delta m_{41}^2 = 1.1\times 10^{-1}$ eV$^2$. These correspond to two distinct points near the $3\sigma$ CL contour of Fig.~3 that was drawn assuming Wilks' theorem holds.

\begin{figure}
    \centering
    \includegraphics[width=0.6\linewidth]{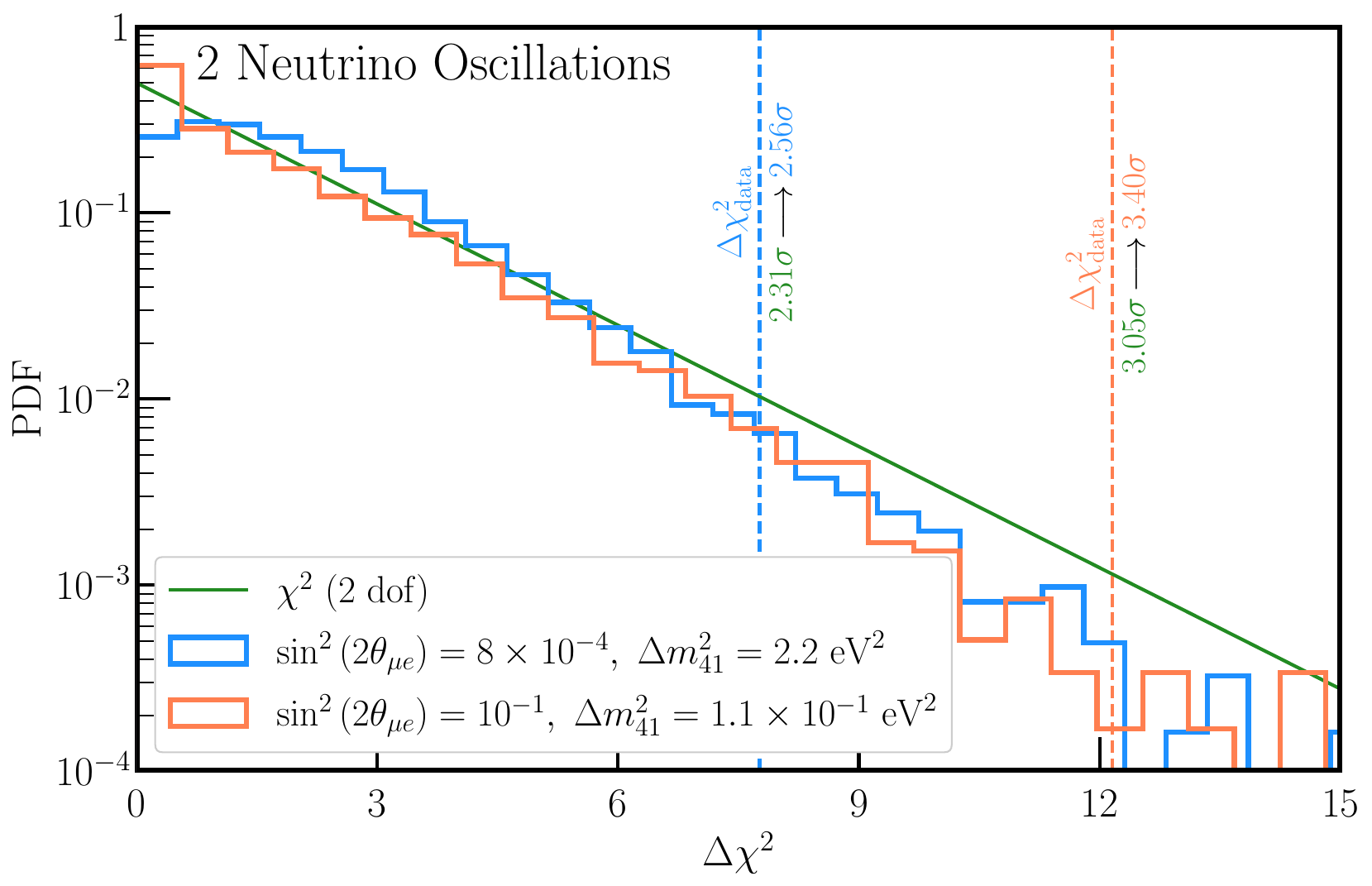}
    \caption{Coverage checks for two test points, blue: $\sin^2\left(2\theta_{\mu e}\right) = 8 \times 10^{-4}$, $\Delta m_{41}^2 = 2.2$ eV$^2$, orange: $\sin^2\left(2\theta_{\mu e}\right) = 10^{-1}$, $\Delta m_{41}^2 = 1.1\times 10^{-1}$ eV$^2$ assuming the two-neutrino oscillation hypothesis (i.e. only $\nu_\mu \to\nu_e$ appearance) for the MicroBooNE-Inclusive search discussed in the main text.\label{fig:Coverage}}
\end{figure}

For each of these two points, we perform a number of pseudoexperiments. For each pseudoexperiment, we generate pseudodata for the MicroBooNE-Inclusive search~\cite{MicroBooNE:2021nxr} assuming that the input parameters are ``true,'' with Poissonian fluctuations about each expectation. This includes the control regions (e.g. $\nu_\mu$ CC samples, etc.) as well. Then, we fit the pseudodata, finding the best-fit point in $\left(\sin^2\left(2\theta_{\mu e}\right), \Delta m_{41}^2\right)$ parameter space, obtaining the difference in the test statistic between that best-fit point and at the assumed-true one. The resulting histograms of this obtained $\Delta\chi^2$ for the pseudoexperiments are shown in Suppl. Fig.~\ref{fig:Coverage} for the two sets of true parameters (a) (blue) and (b) (orange).

When compared against the actual MicroBooNE-inclusive data, the resulting $\Delta\chi^2$ that we obtain, relative to the best-fit point of the full analysis (which corresponds to no-$\nu_e$-appearance), is $7.76$ for (a) and $12.16$ for (b). These benchmarks, as well as the chi-squared distribution for two degrees of freedom, are shown as dashed blue/orange and solid green lines in Suppl. Fig.~\ref{fig:Coverage}, respectively. The p-values obtained with these $\Delta\chi^2$ values, assuming Wilks' theorem holds, are 2.1\% and 0.22\% for points (a) and (b) respectively, which correspond to confidence levels of $2.31\sigma$ and $3.05\sigma$. If we compare these p-values against the fraction of pseudoexperiments for which the obtained $\Delta \chi^2$ is larger than this benchmark, we find updated p-values of 1.0\% and 0.07\% for the two cases. This means that the corrected confidence levels are $2.56\sigma$ and $3.40\sigma$ for cases (a) and (b), implying that using our test statistic assuming Wilks' theorem holds yields (mildly) conservative exclusion regions.

\begin{figure}
    \centering
    \includegraphics[width=0.6\linewidth]{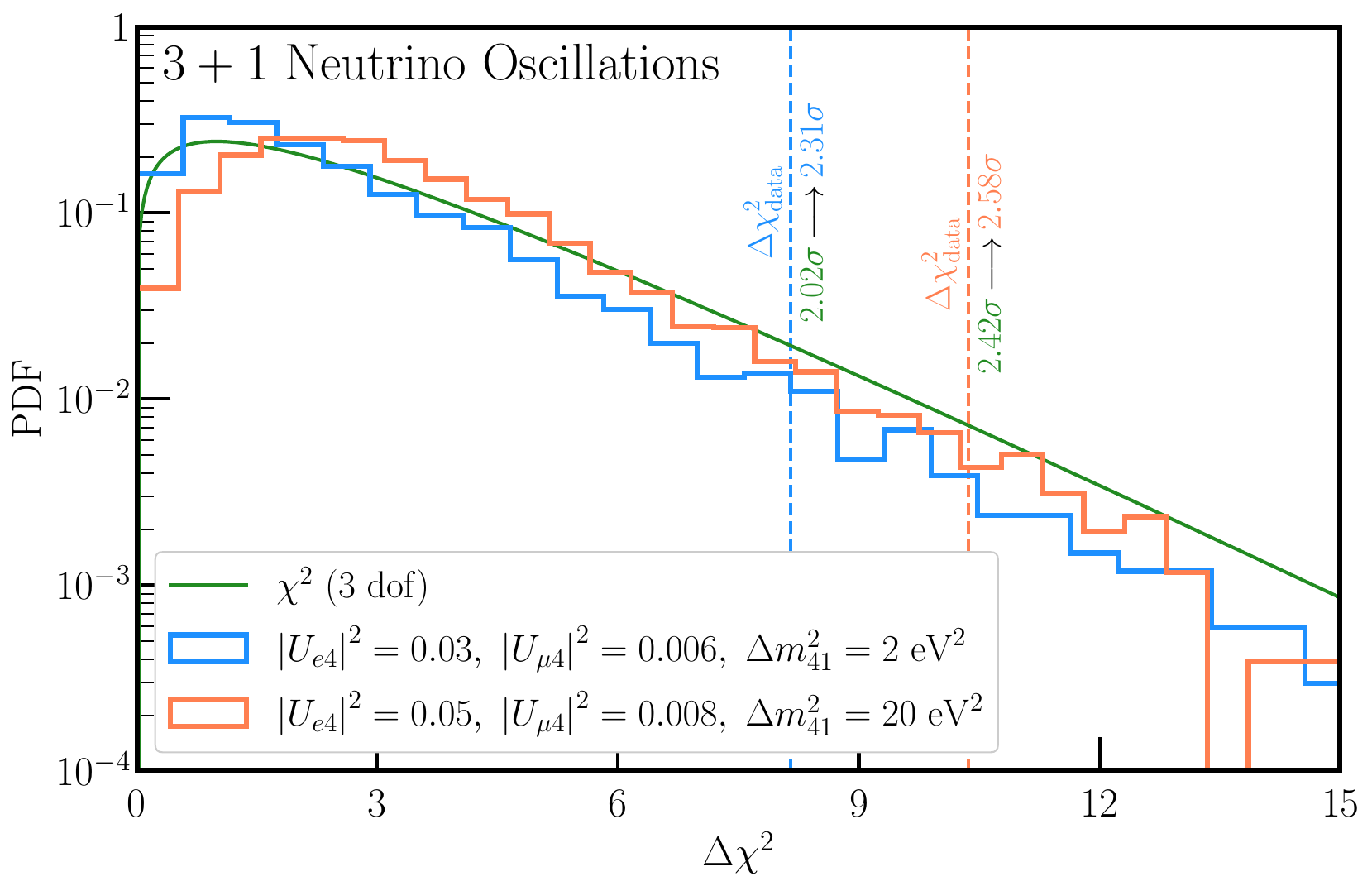}
    \caption{Coverage checks for two test points, left: $\left|U_{e4}\right|^2 = 0.03,$ $\left|U_{\mu4}\right|^2 = 0.006$, $\Delta m_{41}^2 = 2$ eV$^2$, right: $\left|U_{e4}\right|^2 = 0.05,$ $\left|U_{\mu4}\right|^2 = 0.008$, $\Delta m_{41}^2 = 20$ eV$^2$ assuming the full four-neutrino oscillation hypothesis (i.e. including background oscillations of $\nu_\mu$ and $\nu_e$ samples) for the MicroBooNE-Inclusive search discussed in the main text.\label{fig:Coverage3p1}}
\end{figure}
Suppl. Fig.~\ref{fig:Coverage3p1} repeats this process for the full four-neutrino approach discussed in the main text, again, testing the $\Delta\chi^2$ obtained by pseudoexperiments assuming particular points in four-neutrino parameter space. These test points are (a) $\left|U_{e4}\right|^2 = 0.03,$ $\left|U_{\mu4}\right|^2 = 0.006$, $\Delta m_{41}^2 = 2$ eV$^2$ and (b) $\left|U_{e4}\right|^2 = 0.05,$ $\left|U_{\mu4}\right|^2 = 0.008$, $\Delta m_{41}^2 = 20$ eV$^2$. When compared against MicroBooNE-Inclusive data, we obtain $\Delta\chi^2$ relative to the best-fit point at these two points of (a) 8.15 and (b) 10.36. If compared with the expectation of a $\chi^2$ distribution for three parameters (now three with the two distinct mixing angles), this provides confidence levels of $2.02\sigma$ and $2.42\sigma$ when Wilks' theorem is assumed to apply. In contrast, after performing pseudoexperiments, we obtain updated confidence levels of $2.31\sigma$ and $2.58\sigma$, respectively. As with the two-neutrino case surrounding Suppl. Fig.~\ref{fig:Coverage}, we find that our test statistic and the results of Figs.~4 and 5 of the main text yield mildly conservative constraints from the MicroBooNE-Inclusive analysis.

\end{document}